\documentclass{article}
\usepackage{graphicx}
\usepackage{amsmath,amssymb}
\begin{document}
\title{A Look Back at the Ehrenfest Classification.\\
{\Large Translation and commentary of Ehrenfest's 1933 paper introducing the notion of phase transitions of different order}}
\author{Tilman Sauer\thanks{tsauer@uni-mainz.de}}

\maketitle
\begin{abstract}
A translation of Paul Ehrenfest's 1933 paper, entitled ``Phase transitions in the usual and generalized sense, classified according to the singularities of the thermodynamic potential'' is presented. Some historical commentary about the paper's context is also given.
\end{abstract} 
%
%

\section{Introduction}
\label{intro}
The study of systems undergoing either first-order phase transitions or continuous phase transitions has always been in the focus of Wolfhard Janke's research, and many aspects and subtleties of these systems have been elucidated through his work. Phase transitions of higher order have occasionally been studied as well \cite{JankeEtal2006}. But their properties are much more elusive, and models or real world examples of systems undergoing a phase transition of third or even higher order are much harder to find.

On the occasion of reviewing current research on phase transitions and critical phenomena in its whole breadth and variety, it might be interesting to take a look back at the very origin of the distinction of phase transitions of different order. The origin of this distinction can be located quite precisely to one short paper  \cite{Ehrenfest1933} (cited also in Ref.~\cite{JankeEtal2006}), which was published in 1933 by Paul Ehrenfest (1880--1933). Gregg Jaeger has given an excellent historical account of the introduction and evolution of Ehrenfest's classification of phase transitions, and for further details of this history I shall gladly refer to his paper \cite{Jaeger1998}. Based on Jaeger's work, I wish to present here a complete English translation of Ehrenfest's paper along with some explanatory historical commentary.

\section{Some historical context of Ehrenfest's paper}

Going back to the year 1933 takes us back to a time when the very notion of a phase transition was determined almost exclusively by the thermodynamics of transitioning between the solid, liquid, and gaseous phases of homogeneous substances \cite{Brush1976}. Clarifying decades of earlier research, Thomas Andrews in 1869 had investigated the physical characteristics of the critical point, an expression that he had coined, and had established that at the critical point the transition from the liquid to the gaseous state can be continuous \cite{Rowlinson1969}. Phenomena like critical opalescence were known, too, and had been studied theoretically in terms of fluctuations by Marian von Smoluchowski, Albert Einstein, Lenard Ornstein, Frits Zernike and others \cite{EinsteinVol3} but none of these phenomena had been recognized, yet, as a special kind of phase transition. Magnetic systems were being studied, too, and the onset of a ferromagnetic phase at the Curie point was well-known although the study of spin models like those of Ising and Heisenberg were only in its very early beginnings \cite{Niss2005}.  Superconductivity had been discovered in 1911 in Leiden as a sudden and complete loss of electric resistivity of certain metals at liquid helium temperatures. But the discovery of the Meissner-Ochsenfeld effect was reported on only later in that same year 1933 \cite{Dahl1992}.

The immediate occasion for Ehrenfest's introduction of a new class of phase transitions was the discovery of an anomaly in the specific heat of liquid helium at very low temperatures of around 2.19K, the so-called lambda point. The discovery of this anomaly had been made just a few months before in Leiden by Ehrenfest's colleague Willem Hendrik Keesom (1876--1956) and his group.

\subsection{The cryogenic laboratory in Leiden}

The cryogenic laboratory at Leiden founded by Kamerlingh Onnes (1853--1926) had been the first laboratory to succeed in liquifying helium in 1908. In fact, it remained the only place where liquid helium temperatures could be realized for more than a decade, and it was only in 1923 that another laboratory, in Toronto, achieved this capacity with a copy of the Leiden apparatus. Berlin and Charkov joined the list of cryogenic laboratories capable of handling liquid helium in 1925 and 1930, respectively \cite{MatriconWaysand2003}. 

With its longstanding tradition as ``the coldest place on earth'' the Leiden laboratory had dominated low temperature research for several decades and many experimental investigations could only be carried out with the equipment and experience that was available in Leiden. As early as 1885, Kamerlingh Onnes had founded a special publication outlet, named the \emph{Communications from the Laboratory of Physics at the University of Leiden}. Founded as a kind of white paper series for circulation broadly among colleagues at home and abroad, it initially contained English translations or English accounts reporting on the systematic experimentation that was done in Onnes's laboratory. The series ran under this title until 1931, when it was renamed to \emph{Communications from the Kamerlingh Onnes Laboratory of the University of Leiden}. There were also \emph{Supplements} to the \emph{Communications}. In later years, many contributions to the \emph{Communications} or its \emph{Supplements} were published simultaneously as papers in the \emph{Proceedings of the Royal Academy of Sciences} in Amsterdam.

Ehrenfest's 1933 paper is an example of such a contribution that was communicated for publication in the Academy's \emph{Proceedings} and, at the same time, it was included as a \emph{Supplement} in the Laboratory's \emph{Communications}. The paper was presented to the Amsterdam Academy in its meeting of February 25, 1933. At the same meeting, a paper by Willem Keesom reporting ``on the jump in the expansion coefficient of liquid helium in passing the lambda-point'' was presented \cite{Keesom1933}. Keesom's paper was published back to back (pp.~147--152) to Ehrenfest's note (pp.~153--157) in Volume 36 of the \emph{Proceedings}, and it was also included as a \emph{Supplement}. Keesom's paper was originally published in English and carries the number \emph{Supplement 75a}, Ehrenfest's paper appeared in German and carries the number \emph{Supplement 75b}.

\subsection{The ``lambda-point'' of the specific heat of liquid helium}

Keesom had been professor of experimental physics in Leiden since 1923 and a member of the Amsterdam Academy since 1924. As Kamerlingh Onnes's successor as co-director at the cryogenic laboratory he had succeeded to solidify helium in 1926. In his companion paper to Ehrenfest's note, he refers directly to the recent discovery of the lambda-point. That discovery had been made just a few months before and was reported in a first paper \cite{KeesomClusius1932}, co-authored with his collaborator Klaus Clusius (1903--1963), presented for publication by the Academy in its meeting of April 2, 1932 (and published also as \emph{Communication No.~219e}). Somewhat more accurate data had been presented a little later in the Academy session of June 25, 1932, in a paper co-authored with his collaborator and daughter Anna Petronella Keesom \cite{KeesomKeesom1932} (or \emph{Communication No.~221d}).

\begin{figure}
\begin{center}
\includegraphics[scale=0.6]{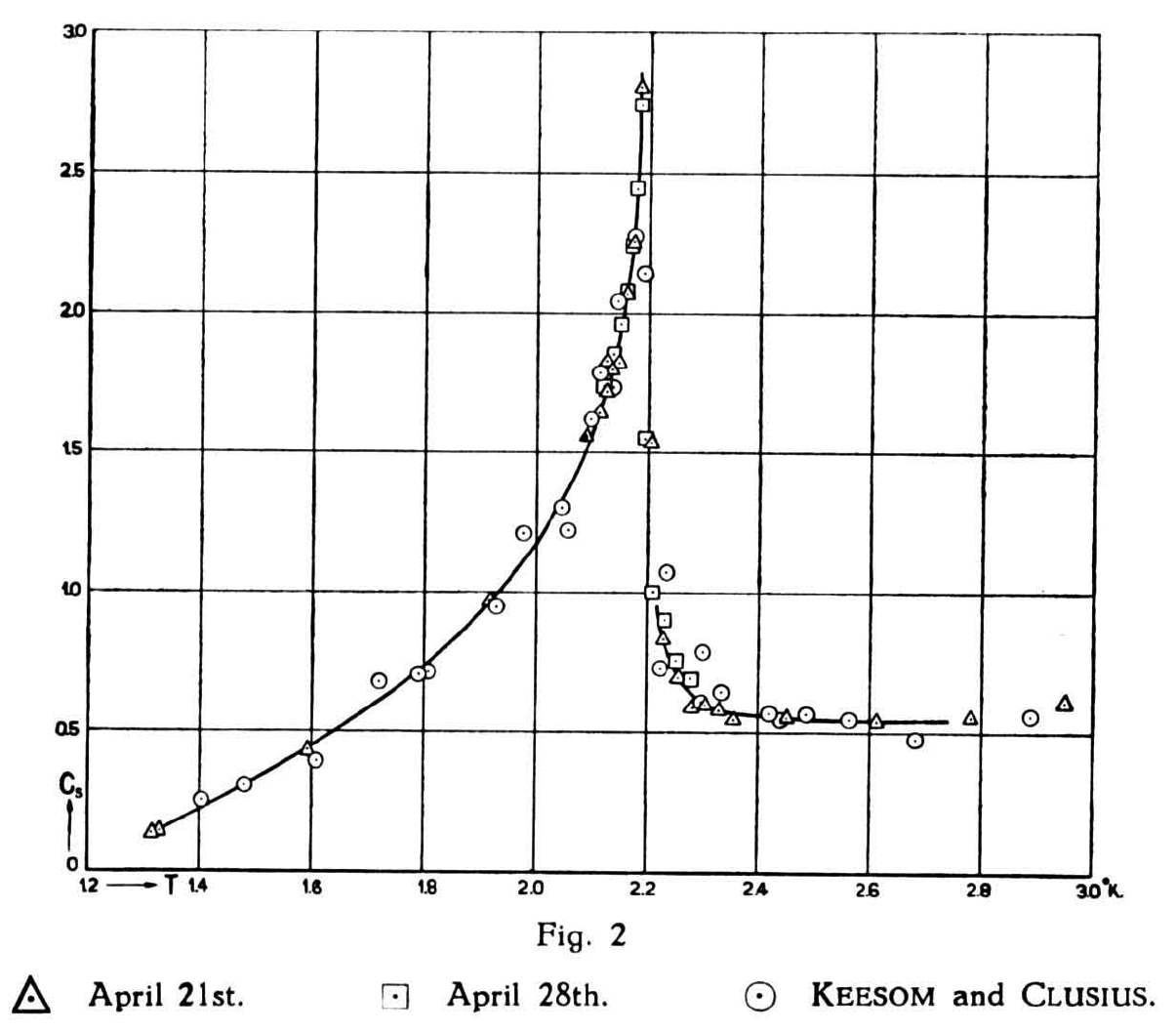}\includegraphics[scale=0.6]{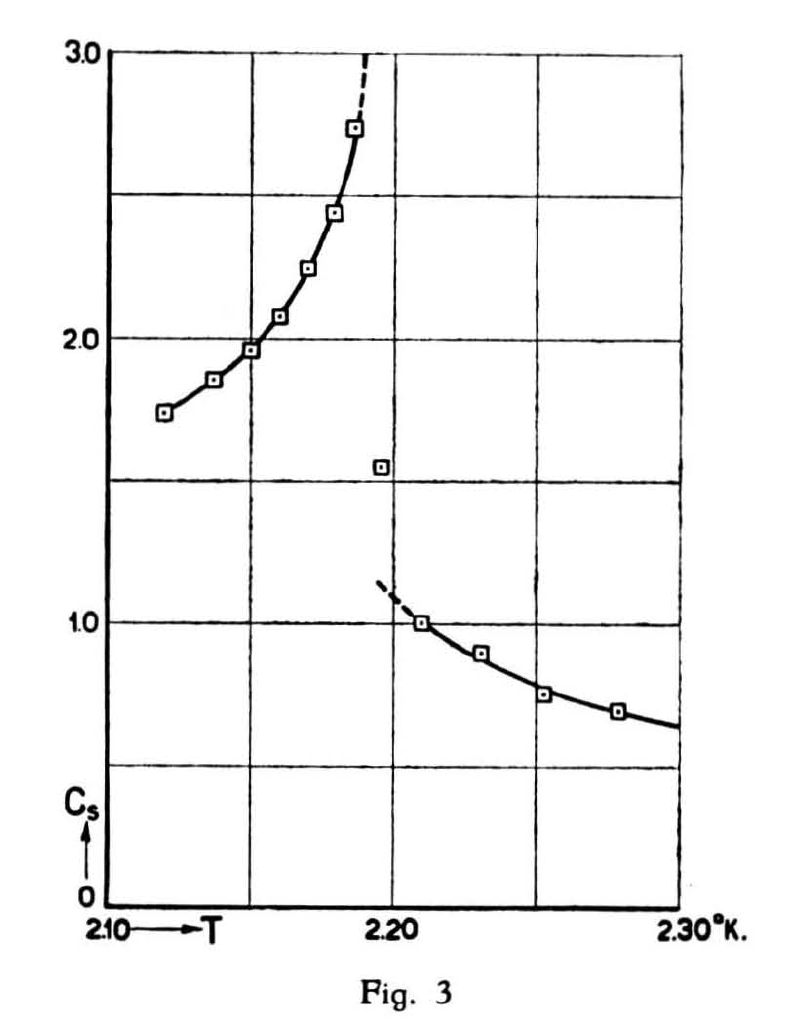}
\caption{Plots of the specific heat data for helium's ``lambda-point,'' as presented in Ref.~\cite[pp.~740, 741]{KeesomKeesom1932}.}
\label{Fig_lambda}       
\end{center}
\end{figure}

Fig.~\ref{Fig_lambda} shows a plot of the specific heat data for the lambda-point as presented in June 1932.
The plot on the left hand side shows the initial data obtained by Keesom and Clusius as circles, together with further data obtained in April 1932 as triangles and squares. The plot on the right hand side shows only the data of April 28 centered around the critical temperature. It is from this second paper, that we also learn about the origin of the name ``lambda-point''. It was introduced by Paul Ehrenfest:
\begin{quote}
``For convenience sake it is desirable to introduce a name for the point at which this jump occurs. According to a suggestion made by Prof.~Ehrenfest we propose to call that point, considering the resemblance of the specific heat curve with the Greek letter $\lambda$, the \emph{lambda-point}.'' \cite[p.~749]{KeesomKeesom1932}
\end{quote}
The name is of some importance, too, because Ehrenfest and his contemporaries interpreted the ``jump'' in the specific heat as a finite discontinuity, not as a (logarithmic) divergence, as we would do now. In fact, a divergence is hard to prove experimentally (for more recent precise measurements of the lambda point under micro gravity conditions, see e.g.\ Ref.~\cite{LipaEtal2003}). We should take note, too, that the discovery in 1932 of the ``lambda-point'' in the specific heat does not mean that the different nature of the two ``phases'' on either side of the specific heat maximum was in any way understood. Indeed, the discovery of superfluidity and its features had to wait for another few years and is credited to Pyotr Kapitza \cite{Kapitza1938} and John F.~Allen and Don Misener \cite{AllenMisener1938} in 1938.

\subsection{Paul Ehrenfest}

Ehrenfest's paper is not a polished account of a well-thought out theory but, on the contrary, it has all the characteristics of a daring, but also somewhat hesitant proposal for conceptual clarification in an ongoing debate. As such it is not untypical of Paul Ehrenfest's style and way of doing physics. 

Ehrenfest grew up in Vienna in a Jewish family with roots in Moravia and studied physics with Ludwig Boltzmann in Vienna \cite{Klein1972}. He also studied for some time in G\"ottingen where he met Felix Klein and David Hilbert. After Boltzmann's death in 1906, Felix Klein asked Ehrenfest to write a review on statistical mechanics for the monumental \emph{Encyclopedia of mathematical sciences}, of which he was one of the main editors. Ehrenfest who had married in 1904 the Russian mathematician Tatyana Afanasyeva took on the job, and in 1911, the two of them delivered a jointly authored review \cite{EhrenfestEhrenfest1911} on \emph{The Conceptual Foundations of the Statistical Approach in Mechanics} for Klein's Encyclopedia. Running to some 90 pages, the review offered an exceptionally clear and lucid account of the basic principles of statistical mechanics with a first-hand knowledge and understanding of Boltzmann's work. The review was translated into English in 1959 \cite{EhrenfestEhrenfest1959}, and is still available as a Dover reprint.

In 1907, Paul and Tatyana Ehrenfest had moved to St.~Petersburg but in 1912, Ehrenfest accepted a call to become Professor of Theoretical Physics in Leiden, the successor of Hendrik Antoon Lorentz. Only a few months before, Ehrenfest and Einstein had met for the first time and had immediately struck up a close friendship. Indeed, Ehrenfest became one of Einstein's closest friends and the extensive correspondence between the two physicists testifies to their many common interests, not only in all aspects of theoretical and experimental physics, but also in political matters as well as about their families, friends, and colleagues. For both of them their Jewish identity also played an important role. In Leiden, Ehrenfest was a very engaged and successful academic teacher who attracted and educated a group of talented physicists. After the First World War, Einstein was appointed, on Ehrenfest's initiative, a special professor at Leiden on a part-time basis, and in the following years, he regularly spent a few weeks each year in Holland, discussing physics with Ehrenfest and his Dutch colleagues.

In his work, like Einstein, Ehrenfest always strove for conceptual clarification of the foundations of physical theories. To the debates of the early quantum theory, he contributed what he called the ``adiabatic principle'' for a sound generalization of Bohr-like quantization rules \cite{PerezEtal2016}. The principle, an alternative to Bohr's correspondence principle, asserts that quantization postulates that hold for one mechanical system may be transformed to another mechanical system if an adiabatic transformation between the classical mechanical systems can be found. When Otto Stern and Walther Gerlach published their famous experimental verification of space quantization in 1922, Einstein and Ehrenfest were among the first to realize the significance of the result \cite{SchmidtBoeckingEtal2016SternGerlach}. They immediately published a theoretical analysis showing that the Stern-Gerlach findings could not be explained on the grounds of classical physics and they, indeed, anticipated, in a sense, the problematic of the quantum measurement problem \cite{UnnaSauer2013}.

As his christening of the anomaly of the specific heat curve as a ``lambda-point'' already showed, Ehrenfest had a penchant and talent for the creation of witty and fitting terms and phrases. When their second daughter, who was named after her mother Tatyana, developed an interest in mathematics as well, like her mother, he referred to her as Tatyana$^\prime$ (``Tatyana prime''). And in the late twenties, when Ehrenfest, who had an excellent mathematical training, found it difficult to understand the modern formulation of quantum mechanics and especially the role of group theory in it, he coined the word ``Gruppenpest'' (plague of group theory) to refer to the predominance of formal mathematical methods over conceptual understanding \cite[p.~63]{Schneider2011}. Tragically, Ehrenfest found it harder and harder in the late twenties and early thirties to follow up with modern developments in theoretical physics, in particular with modern quantum mechanics which he found messy and difficult to understand in their conceptual foundations. On 25 September 1933, a mere half year after his paper on the classification of phase transition, Ehrenfest committed suicide. Einstein wrote a moving obituary for his close friend \cite{Einstein1934}.

\subsection{The argument of the paper}

The paper on the classification of phase transitions is Ehrenfest's very last publication, and for this reason alone it deserves to be remembered. Not at all written as a legacy paper or a summarizing review of some long and laborious work, it rather throws out an idea that opened up a new way of looking at phase transitions. 

The paper itself conveys a simple point. Ehrenfest starts with a reference to the recent discovery of the lambda-point in the specific heat of helium by Keesom and his co-workers, citing Refs.~\cite{KeesomClusius1932} and \cite{KeesomKeesom1932} as well as \cite{KeesomClusius1931}. Argueing for the interpretation of the lambda-point as a phase transition, he also points out what the dissimilarity with known phase transitions was, namely the absence of any latent heat or of a change in volume.

Almost as if he felt he had to justify his publishing a very preliminary and unpolished idea, Ehrenfest refers to recent discussions in Leiden and mentions that his colleague Keesom had suggested that he publish his idea. His argument laid out in the following pages then is this. If the lambda-point indicates a phase transition, it cannot be a usual one since there is no discontinuity in the entropy nor in the volume. But there is a discontinuity, or so he would interpret it, in the specific heat. It seems that, at this point, Ehrenfest might have remembered a classic argument going back to Gibbs \cite{Gibbs1873} of deriving the Clausius-Clapeyron equation from a consideration of the thermodynamic potential and its derivatives. Working in a representation with energy and entropy as independent variables, Gibbs showed how to obtain the Clausius-Clapeyron relation from a graphical representation of the thermodynamical potential \cite[pp.~387--388]{Gibbs1873}. Transferring the very same argument for the new case at hand, and working in a $p$-$T$-representation, Ehrenfest now derives the analogue for the Clausius-Clapeyron equation for the case of continuous first but discontinuous second derivatives of the thermodynamic potential. This derivation carries weight because Ehrenfest arrives at the very same relation that Keesom in Ref.\cite{Keesom1933} had obtained by considering a thermodynamic cyclic process and found confirmed by his experimental data.

After giving his brief derivation of the new Clausius-Clapeyron relation, Ehrenfest ends by raising a number of open research questions to follow up on. In these, he points to different physical phenomena that might be classified successfully by his new scheme, to wit some recent experimental work by Franz Simon (1893--1956) on similar ``bumps'' in specific heat measurements of ammonium chloride at a temperature of ca. 242.6 K and similar phenomena \cite{Simon1922} as well as attempts to explain these in terms of quantum excitations \cite{Simon1926}. He also refers to the phenomena of ferromagnetism and superconductivity and, indeed, in an addendum at proof stage, he indicates that his proposal was taken up by colleagues and applied to the case of superconductive phase transitions. He also calls for an ``\emph{essential} kinetic interpretation'' of higher-order phase transitions, and he points out that it seems that for second-order phase transitions it seems to be impossible to have two phases coexistent. He adds:
\begin{quote}
``I would wish very much that I were capable of formulating and understanding this characteristic difference with respect to ``usual'' transitions in a better way.''
\end{quote}

\section{Concluding remarks}

A discussion of the immediate as well as long term reception of Ehrenfest's paper can be found in Jaeger's paper \cite{Jaeger1998} who also points out a curious irony. The very phenomenon that gave rise to Ehrenfest's proposal of classifying phase transitions according to the derivatives of the thermodynamic potential, i.e. the lambda-point transition was later understood to fall outside of this classification. Indeed, the transition from He I to superfluid He II at the lambda-point displays a logarithmic divergence of the specific heat rather than a simple discontinuity. Later authors therefore realized the necessity to extend or modify Ehrenfest's classification scheme if one wants to hold on to it at all.

Nevertheless, Ehrenfest's paper remains important for several reasons. First, it introduced a new kind of phase transition and it gave a specific meaning to the difference between first order phase transitions and continuous ones. Second, its implicit introduction of phase transitions of third and higher order has proven to be a fruitful idea for further investigations into the nature of phase transitions. And finally, it remains a remarkable example of conceptual innovation that arises from combining a purely mathematical framework with a penetrating understanding of physical phenomena in a new and ground breaking way.

\section*{Phase transitions in the usual and generalized sense, classified according to the singularities of the thermodynamic potential\footnote{[The following is an English translation of Ehrenfest's paper \cite{Ehrenfest1933}.]}}

\begin{center}

\noindent by P.~Ehrenfest

\bigskip

\noindent (Communicated at the meeting of February 25, 1933.)

\bigskip

\emph{Summary:}
\end{center}
The measurements by Keesom and collaborators of the characteristic behavior of the specific heat of liquid helium and also of that of the superconductors suggest a certain generalization of the concept of a phase transition. Discontinuity curves of different order on the surface of the thermodynamic potential turn into transition curves for the ``transitions of first, second, and higher order''. For the usual phase transitions we obtain the Clapeyron equation between the jumps of the first differential quotients of the thermodynamic potential, i.e. between $S''-S'$ and $v''-v'$. For those of second order we obtain analogous equations between the jumps of the specific heat and the jumps of $\frac{\partial v}{\partial T}$ and $\frac{\partial v}{\partial p}$.

\bigskip

The anomaly in the behavior of the specific heat of liquid helium, which was discovered by Keesom and collaborators%
\footnote{W.H.~Keesom and K.~Clusius. These Proceedings {\bf 35}, 307, 1932. Comm.\ Leiden N$^o$.~219. W.H.Keesom and Miss~A.P.~Keesom. These Proceedings {\bf 35}, 736, 1932. Comm.\ Leiden N$^o$.~221d.}
and which appears to be a discontinuity at the currently available experimental precision,  as well as the shift of the ``lambda point'' under pressure along a ``lambda-point-curve'' in the $p$-$T$-plane studied by him together with Clusius%
\footnote{W.H.~Keesom and K.~Clusius. These Proceedings {\bf 34}, 605, 1931. Comm.\ Leiden.\ N$^o$.~216b.}
justify the interpretation of this curve as a transition curve between two modifications of liquid helium: He I and He II, i.e.~as a \emph{$p$,~$T$-transformation curve between two (liquid) phases}.---It also fits the interpretation that Keesom%
\footnote{Proceedings of this meeting. Comm.\ Leiden Suppl.\ N$^o$.~75a.}
was able to derive a relation between the jump of the specific heat, on the one hand, and the jump of the thermal expansion coefficient, on the other hand, by means of considering a cyclic process and that he was able to establish a satisfactory agreement with the measurements.

Because of the suggestive \emph{similarity} with a phase transition it is all the more interesting to take a closer look at the characteristic \emph{dissimilarity}, too---namely the absence of an entropy difference (latent heat) and of a volume difference between these two phases.

In a very instructive discussion, in which Keesom directed my attention to these circumstances, it became evident that one can very conveniently formulate the peculiar \emph{generalization of the concept of a phase transition} which is suggested by the discovery of the lambda-point-curve in the language of the thermodynamic potential of the zeta function $Z(T,p)$. I believe to be justified in following Keesoms suggestion to publish these remarks because an analogous kind of formulation will likely prove to be convenient for the behavior of superconductors at the jump temperature and of the ferromagnets at the Curie temperature.

\section*{\S~1. Singular curves of different order on the $Z(T,p)$-surface}

I begin by recalling the following equations
\begin{equation}
\frac{\partial Z}{\partial T} =-S \label{eq1}
\end{equation}
\begin{equation}
\frac{\partial Z}{\partial p} =v \label{eq2}
\end{equation}
\begin{equation}
\frac{\partial^2 Z}{\partial T^2} = -\frac{\partial S}{\partial T} = -\frac{C}{T} \label{eq3}
\end{equation}
\begin{equation}
\frac{\partial^2 Z}{\partial p} =\frac{\partial v}{\partial p} \label{eq4} 
\end{equation}
\begin{equation}
\frac{\partial^2 Z}{\partial T\partial p} = -\frac{\partial S}{\partial p} = -\frac{\partial v}{\partial T},
\label{eq5}
\end{equation}
where $Z(T,p)$ denotes the thermodynamic potential, $S$ the entropy, $c$ the specific heat at constant $p$. In general, $Z(T,p)$ is continuous with all lower differential quotients. But now let us look at a piece of a ``transition curve'' in the $p,T$-plane, where discontinuities are found that will be discussed below.
\begin{center}
\includegraphics[scale=0.7]{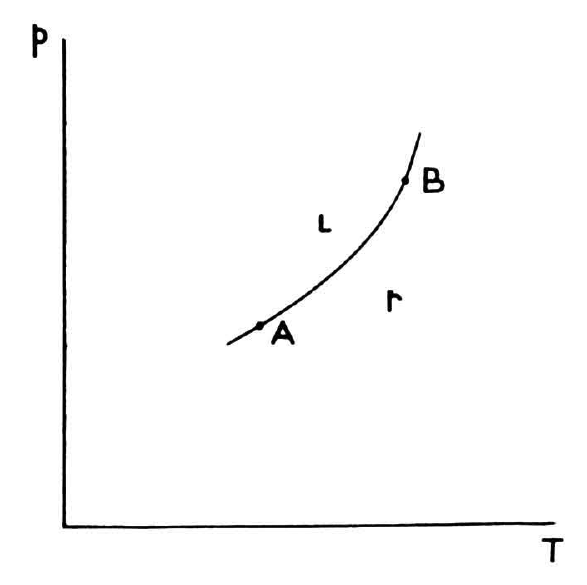}
\end{center}

\emph{In any case}, we can disregard the possibility of a jump of $Z$ itself. This means that in any case the relation 
\begin{equation}
Z_{\rm r}-Z_{\rm l} = ((Z)) = 0. \tag{I}
\label{eqI}
\end{equation}
should hold along the entire transition curve $A$, $B$.
This is because a jump of $Z$ would mean that volume and entropy would become infinite due to \ref{eq1}, \ref{eq2}.

On the other hand, one would admit the possibility that:
\begin{align}
\left( \frac{\partial Z}{\partial T}\right)_{\rm r} - \left( \frac{\partial Z}{\partial T}\right)_{\rm l}
&= \left(\left( \frac{\partial Z}{\partial T}\right)\right) \neq 0 \\
\left( \frac{\partial Z}{\partial p}\right)_{\rm r} - \left( \frac{\partial Z}{\partial p}\right)_{\rm l}
&= \left(\left( \frac{\partial Z}{\partial p}\right)\right) \neq 0 
\end{align}
i.e.\ that the $Z(T,p)$-surface over the curve $A$, $B$ is bent.

This ``discontinuity of first order'' is given with the usual phase transitions of first order, since there we have (see \ref{eq1}, \ref{eq2}):
\begin{align}
\left(\left( \frac{\partial Z}{\partial T}\right)\right) &= ((S)) = \frac{Q}{T} \qquad
\text{(Q the latent heat)} \tag{7a} \label{eq7a} \\
\left(\left( \frac{\partial Z}{\partial p}\right)\right) &= ((v)) = v_{\rm r}-v_{\rm l} \qquad
\text{(volume difference of unit mass in both phases)}
\tag{8a} \label{eq8a}
\end{align}
With a \emph{discontinuity of second order}, however, we shall have
\begin{equation}
\text{(I)} \quad ((Z))=0\qquad 
\text{(II)} \left(\left( \frac{\partial Z}{\partial T}\right)\right) = 0 \qquad
\text{(III)} \left(\left( \frac{\partial Z}{\partial p}\right)\right) = 0
\end{equation}
and only then
\begin{align}
\left(\left( \frac{\partial^2 Z}{\partial T^2}\right)\right) &= 
- \frac{((c))}{T} \neq 0 \label{eq9}\\
\left(\left( \frac{\partial^2 Z}{\partial p^2}\right)\right) &= 
\left(\left( \frac{\partial v}{\partial p}\right)\right) \neq 0\\
\left(\left( \frac{\partial^2 Z}{\partial T\partial p}\right)\right) &=
\left(\left( \frac{\partial v}{\partial T}\right)\right) =
- \left(\left( \frac{\partial S}{\partial p}\right)\right)\neq 0
\end{align}
(cp.~\ref{eq3}, \ref{eq4}, \ref{eq5}). The relation (\ref{eq9}) shows that such a discontinuity of second order is given just with Keesom's lambda-point-curve where the specific heat is discontinuous but $Q=0$ and $((v))=0$ still hold.

\section*{\S~2.\ Clapeyron's equation and the analogous relations in the case of phase transitions of higher order}

If for any quantity $G$ it can be confirmed that it does \emph{not} display a jump at the ``transformation curve'' $A$, $B$, i.e.\ along the entire curve
\begin{equation}
((G))=0, \tag{A} \label{eqA}
\end{equation}
holds, then we have
\begin{equation}
Dp \left(\left( \frac{\partial G}{\partial p}\right)\right) + 
DT \left(\left( \frac{\partial G}{\partial T}\right)\right) = 0,
\tag{B} \label{eqB}
\end{equation}
or:
\begin{equation}
\frac{Dp}{DT} = 
-\frac{\left(\left( \frac{\partial G}{\partial T}\right)\right)}{\left(\left( \frac{\partial G}{\partial p}\right)\right)},
\tag{C} \label{eqC}
\end{equation}
where the $Dp$, $DT$ written with capitals denote taking the differentials \emph{along} the transformation curve. Therefore it follows from (\ref{eqI}) because of (\ref{eqC}, \ref{eq7a}, \ref{eq8a})
\begin{equation}
\frac{Dp}{DT} = \frac{Q}{T(v_{\rm r}-v_{\rm l})} \tag{D}
\label{eqD}
\end{equation}

In the case of a discontinuity of \emph{first} order, this is the equation of Clapeyron. In the case of a discontinuity of \emph{second} order the right hand side degenerates into $0/0$.---On the other hand, because of (II) and (III) we have: along the entire transformation curve $A$, $B$:
\begin{equation*}
(II') \quad ((S)) = 0 \qquad (III') \quad ((v)) = 0
\end{equation*}
that is the quantities $S$ and $v$ here show the behavior (\ref{eqA}). There one has in this case because of (\ref{eqC}):
\begin{align}
\frac{Dp}{DT} &= - \frac{\left(\left( \frac{\partial S}{\partial T}\right)\right)}{\left(\left( \frac{\partial S}{\partial p}\right)\right)} = \frac{((c))}{T\left(\left(\frac{\partial v}{\partial T}\right)\right)} \tag{E} \label{eqE} \\
\frac{Dp}{DT} &= - \frac{\left(\left( \frac{\partial v}{\partial T}\right)\right)}{\left(\left( \frac{\partial v}{\partial p}\right)\right)} \tag{F}\label{eqF}
\end{align}
(cp.~\ref{eq3}, \ref{eq5}). (\ref{eqE}) is the relation that was derived and experimentally tested by Keesom. From (\ref{eqE}) and (\ref{eqF}) it also follows that:
\begin{equation}
((c)) = -T \frac{\left(\left( \frac{\partial v}{\partial T}\right)\right)^2}{\left(\left( \frac{\partial v}{\partial p}\right)\right)} \tag{G} \label{eqG}
\end{equation}

\section*{\S~3.\ Some remarks}

a. Our considerations only refer to the occurrence of \emph{discontinuities} in the specific heat.%
\footnote{I cannot judge whether one could approximately idealize in this sense also e.g.\ the anomalies discovered by F.~Simon, Ann.\ d.\ Phys.\ (4) 68, 241.} 
We did not comment on the possible occurrence of ``bumps''%
\footnote{F.~Simon. Berlin Sitz. Ber. 1926, p.~477.}
in their functional behavior.  One would want to treat these as something like an unsharp phase transition.

b. Although we believe it to be useful to talk about ``transformation of one phase into another one'' also in the case of discontinuities of second order, it does not seem to be possible to have both phases ``spatially coexistent'' in equilibrium  in this case. I would wish very much that I were capable of formulating and understanding this characteristic difference with respect to ``usual'' phase transformations in a better way.

c. Here one feels especially well the kinship with the transformation into the superconducting state and into the ferromagnetic state. In the latter case, by the way, we don't seem to be dealing with a $t$, $H$-discontinuity \emph{curve} but only with a \emph{point}: $H=0$, $T=t_{\rm c}$.

d. Searching for the \emph{essential} kinetic interpretation of the above-discussed discontinuities, one gets to conjecture that it is associated with the following event: the distribution of hypersurfaces of constant total energy in the ``\emph{Gamma}-phase-space'' of the system has to be of such a kind that the volume contained by consecutive energy surfaces $V(E)$ displays an unusally high value of $dV/dE$ for a certain energy value $E_0$.

\emph{Addendum at proof stage}.--- Dr.~A.J.Rutgers made the following remark in a discussion regarding the application to superconductors: If one substitutes in (\ref{eqE}) and (\ref{eqF}) the quantities $p$ and $v$ by the magnetic field strength and the magnetization, and if one multiplies the resulting equations with another, one obtains a relation between: \emph{on the one hand} $DH/DT$, the shift of the jump temperature with the applied field (see the investigations by W.J.de Haas and his collaborators). For tin there appears to be good agreement. In addition, Dr.~P.M.van~Alphen points out that the very large values of $DH/DT$ for certain alloys (e.g.\ for Bi$_5$Tl$_3$---see J.~Voogd, diss. Leiden, 1931) lead us to expect particularly high $c$-jumps! Dr.~Rutgers hopes to report elsewhere on all this in more detail, as soon as doubts about the exact meaning of the quantity that is analogous to $\partial v/\partial p$ will have been resolved.

\end{document}